\documentclass[12pt]{article}
\usepackage[a4paper, left=1in, right=1in, top=1in, bottom=1in]{geometry}
\usepackage{amsmath}
\usepackage{mathrsfs}
\usepackage{dutchcal}
\usepackage{amssymb}
\usepackage{amsfonts}
\usepackage{graphicx}
\usepackage{subcaption}
\usepackage{hyperref}
\usepackage{geometry}
\usepackage{float}
\usepackage{tikz}
\usepackage{tikz-3dplot} 
\usepackage{authblk}
\usepackage{ulem}
\usepackage[numbers,sort&compress]{natbib}

\title{\bfseries Quantum Entanglement Autodistillation in Baryon Pair Decays}

\author{Hai-Long Feng, Hao Tang, Wu-zhong Guo\thanks{Email: \texttt{wuzhong@hust.edu.cn}}, and Qin Qin\thanks{Email: \texttt{qqin@hust.edu.cn}}}
\affil{\small School of Physics, Huazhong University of Science and Technology, Wuhan 430074, China}

\date{\today}

\begin{document}

\maketitle

\begin{abstract}
We study the spin-entangled mixed state of a spin-1/2 baryon–antibaryon pair produced in the process \(e^+e^- \to J/\psi, \psi' \to \mathcal{B}\bar{\mathcal{B}}\). We show that the spin entanglement of the system can increase following the decays \(\mathcal{B} \to \mathcal{b} + M\) and \(\bar{\mathcal{B}} \to \bar{\mathcal{b}} + \bar{M}\), where \(\mathcal{b}, \bar{\mathcal{b}}\) are spin-1/2 baryons and \(M, \bar{M}\) are spin-0 mesons. This phenomenon, known as entanglement autodistillation, manifests as a probabilistic amplification of entanglement during the decay process. We analyze the underlying mechanism and show that this phenomenon depends only on the initial state of the \(\mathcal{B}\bar{\mathcal{B}}\) system and the decay parameter \(\alpha_D\), but not on the decay parameter \(\phi_D\). This effect only arises when parity is violated in the decay, when \(\alpha_D \neq 0\).

\end{abstract}

\section{Introduction}
Quantum entanglement is one of the defining features of quantum mechanics that distinguishes it from classical theories. It has found widespread applications in fields such as metrology, cryptography, quantum information, and quantum computing. Entanglement has been experimentally observed across different physical systems and length scales, ranging from the microscopic to the macroscopic. As a fundamental concept in quantum mechanics, it plays a crucial role in both theoretical research and practical applications\cite{Horodecki:2009zz}.  

The application of quantum entanglement concepts to particle physics has become a growing area of interest in recent years\cite{Rosenfeld:2017rka, Wu:2024asu, Li:2006fy, Ehataht:2023zzt, Fabbrichesi:2024rec, Guo:2024jch, Fabbrichesi:2023idl, Chen:2024syv, Geng:2023ffc, Bernal:2024xhm, Chen:2024drt, Zhang:2025mmm, Afik:2025ejh, Han:2025ewp, Han:2024ugl, Bi:2024tlq, Bi:2023uop, Bai:2024omg, Bai:2023hrz, Bai:2023tey, Bai:2022hfv, Du:2024sly, Cheng:2025cuv,  Morales:2023gow, Morales:2024jhj}. A recent study\cite{ATLAS:2023fsd} reported the observation of quantum entanglement in top-quark pairs at the LHC using the ATLAS detector. This represents the highest energy scale at which entanglement has been observed to date. The top quark, the heaviest known elementary particle, has an extremely short lifetime (\(10^{-25}s\)), significantly shorter than the hadronization timescale. Theoretical studies on top-quark entanglement have made substantial progress \cite{Parke:1996pr, Bernreuther:2004jv, Baumgart:2012ay, Mahlon:2010gw}, and experimental observations hold significant implications: on one hand, they provide a test of fundamental quantum entanglement concepts at high energy and ultrashort timescales; on the other hand, entanglement may offer novel methods for probing particle physics. This research opens new perspectives and opportunities at the intersection of quantum information and quantum field theory.

Recent studies \cite{Aguilar-Saavedra:2024fig, Aguilar-Saavedra:2024hwd, Aguilar-Saavedra:2023lwb, Aguilar-Saavedra:2023hss} have shown that particle decay processes can significantly change a system’s entanglement, potentially leading to an anomalous phenomenon known as “entanglement amplification”. This process, referred to as “autodistillation”, suggests that entanglement can spontaneously increase during the system’s evolution, although with some probability of failure. According to quantum information principles, entanglement cannot increase through local operations and classical communication (LOCC). However, under suitable local manipulations, it may increase probabilistically. These operations, known as stochastic LOCC (SLOCC) \cite{Bennett:2000fte}, form the theoretical basis of autodistillation. The transformation of entanglement in two-qubit mixed states under SLOCC operations has been studied in Ref.~\cite{Verstraete_2001}, where the so-called ``filtering operations'' are identified as SLOCC transformations.

In this work we study the spin-entangled system for a baryon and antibaryon pair, which can be described by an entangled mixed state. To characterize the entanglement, we adopt negativity and concurrence as quantitative measures. It is found the spin entanglement can increase after the decay of baryon and antibaryon, demonstrating the phenomenon of autodistillation in the hadron system. In our analysis, we exclude the influence of the accompanying mesons since they are pseudoscalars. The pseudoscalar nature of the accompanying mesons ensures that they do not carry spin information and hence do not interfere with the entanglement. This feature distinguishes our setup from that in Ref.~\cite{Aguilar-Saavedra:2024fig}, where the decay \(t \to Wb\) involves a spin-1 \(W\) boson, introducing additional spin degrees of freedom. Furthermore, our proposed scenario can be experimentally accessed at \(e^+e^-\) colliders. Employing the decay parameters in the involved baryon pair decays measured by experiments, we verify the observation of the phenomena of quantum entanglement autodistillation in the baryon pair decays. We suggest the experiments to directly measure the momentum correlation of the decay products.

The remainder of this paper is organized as follows.  
In Section~\ref{sec:ent_measure}, we introduce the entanglement measures—negativity and concurrence—used throughout our analysis.  
Section~\ref{sec:process} details the specific production and decay processes of baryon–antibaryon pairs, along with their density matrix formalism.  
Our main results demonstrating entanglement autodistillation are presented in Section~\ref{sec:results}, with separate subsections analyzing different decay channels.  
In Section~\ref{sec:reconstruction}, we describe how to reconstruct the spin configuration of the mother particles using the angular distributions of the daughter particles' momenta. This technique enables the detection of entanglement autodistillation through direct measurements of momentum correlations.  
Section~\ref{sec:mechanism} explores the underlying mechanism of entanglement autodistillation and shows that the phenomenon depends solely on the initial state and the decay parameter \(\alpha_D\), while remaining unaffected by the parameter \(\phi_D\).  
Finally, we conclude with a discussion on the implications of our findings and propose that the autodistillation effect could be experimentally tested at future \(e^+e^-\) colliders.

\section{Entanglement Measures}\label{sec:ent_measure}
Entanglement is tantamount to correlations that cannot be described in terms of classical probabilities.
In a pure state, no classical correlations exist, allowing us to constructively define certain entanglement measures, such as entanglement entropy. However, in a mixed state, classical correlations are present, making it challenging to define entanglement measures, as one must distinguish the quantum correlations from classical ones \cite{Mintert2009}. For mixed states, {\it negativity}\cite{Vidal_2002} and {\it concurrence}\cite{Wootters_1998} are considered to be two useful entanglement measures. Since the system under consideration in this work is a mixed state, we adopt these two measures to characterize the amount of entanglement. We will briefly introduce them in this section.

For a bipartite quantum system with Hilbert space \(\mathcal{H} = \mathcal{H}_1 \otimes \mathcal{H}_2\), entanglement can be characterized by the \textit{negativity}, which is derived from the partial transpose of the density matrix \(\rho\). A separable (i.e., unentangled) mixed state takes the form  
\[
\rho = \sum_i p_i\, \rho_i^{(1)} \otimes \rho_i^{(2)},
\]  
where \(\{p_i\}\) is a probability distribution and \(\rho_i^{(1)}, \rho_i^{(2)}\) are local density matrices.

The \textit{partial transpose} operation transposes one of the two subsystems in a fixed basis. For separable states, the partially transposed density matrix \(\rho^{\text{pt}}\) remains positive semi-definite, and thus all its eigenvalues are non-negative. However, for entangled states, the partial transpose may contain negative eigenvalues, which signals the presence of entanglement. This observation motivates the definition of negativity as an entanglement measure as \cite{Vidal_2002}
\[
\mathcal{N}(\rho)=\frac{\sum_i|\lambda_i|-1}{2}.
\]
Here, \( \mathcal{N}(\rho) \) represents the negativity of \( \rho \), and \( \lambda_i \) denotes the eigenvalues of the partially transposed density matrix \( \rho^{\text{pt}} \). 
The partial transpose does not influence the trace of the density matrix, so if some of the eigenvalues \(\lambda_i\) of \(\rho^{\text{pt}}\) are negative, then \(\sum_i|\lambda_i|\) must be bigger than one. Negativity is fully reliable for \(2\times2\) systems~\cite{Mintert2009}, including the baryon pair systems investigated in this study.

The concurrence of a mixed state in a bipartite two-level system can be defined as follows. An auxiliary matrix is introduced:  
\[
R=\rho(\sigma_y\otimes\sigma_y)\rho^*(\sigma_y\otimes\sigma_y),
\]  
where \( \rho^* \) denotes the matrix whose entries are the complex conjugates of those in \( \rho \). The eigenvalues of \( R \) are then computed, and their square roots are denoted as \( r_i \), \( i=1,2,3,4 \). For convenience, \( r_1 \) is chosen as the largest one. The concurrence of the bipartite two-level system is then defined as \cite{Wootters_1998}:  
\[
\mathscr{C}(\rho) \equiv \max(r_1 - \sum_{i>1} r_i, 0).
\]

For a pure state \( |\psi\rangle \), this expression reduces to \( \mathscr{C}(|\psi\rangle) = 2 |\det C| \), where \( C \) is the coefficient matrix of \( |\psi\rangle \) in the standard basis. For mixed states, the concurrence can also be derived via the \textit{convex roof construction}~\cite{Mintert2009}, guaranteeing consistency with the pure-state definition.

\section{Specific Process}\label{sec:process}
We begin by describing the production and decay processes of spin-entangled baryon-antibaryon pairs. The system under study involves two sequential processes: first, the production of a baryon-antibaryon ($\mathcal{B}\bar{\mathcal{B}}$) pair from $e^+e^-$ annihilation through vector meson resonances ($J/\psi$ or $\psi'$), followed by the weak decays of both baryons into daughter particles.

The production process $e^+e^- \to J/\psi,\psi' \to \mathcal{B}\bar{\mathcal{B}}$ creates a spin-entangled $\mathcal{B}\bar{\mathcal{B}}$ pair, where $\mathcal{B}$ represents a spin-$\frac{1}{2}$ baryon and \(\bar{\mathcal{B}}\) represents a spin-$\frac{1}{2}$ antibaryon. Subsequently, each baryon undergoes weak decay:
\begin{itemize}
    \item $\mathcal{B} \to \mathcal{b} + M_0$ (baryon decay)
    \item $\bar{\mathcal{B}} \to \bar{\mathcal{b}} + \bar{M}_0$ (antibaryon decay)
\end{itemize}
where $\mathcal{b}$ ($\bar{\mathcal{b}}$) is a daughter spin-$\frac{1}{2}$ baryon (antibaryon) and $M_0$ ($\bar{M}_0$) is a spin-0 meson. The entanglement initially present in the $\mathcal{B}\bar{\mathcal{B}}$ system may be modified through these decay processes.

The complete process can be characterized by kinematic variables $\xi = (\theta_1, \theta_\mathcal{b}, \phi_\mathcal{b}, \theta_{\bar{\mathcal{b}}}, \phi_{\bar{\mathcal{b}}})$, where $\theta_1$ represents the scattering angle of the \(\mathcal{B}\) baryon with respect to the positron beam in the center-of-mass (CM) frame. The remaining four angles, \(\theta_\mathcal{b}, \phi_\mathcal{b}, \theta_{\bar{\mathcal{b}}}, \phi_{\bar{\mathcal{b}}}\), describe the orientation of the \(\mathcal{b}\) and \(\bar{\mathcal{b}}\) baryons in their respective helicity frames, denoted as \(\mathcal{R}_\mathcal{B}\) and \(\mathcal{R}_{\bar{\mathcal{B}}}\), where the \(\mathcal{B}\) and \(\bar{\mathcal{B}}\) baryons are at rest.  

\begin{figure}[H]
    \centering
    \begin{tikzpicture}
    \tdplotsetmaincoords{60}{10}  
    \begin{scope}[tdplot_main_coords] 

        \filldraw[gray!20, opacity=0.5, draw=black] 
            (-3,-2,0) -- (3,-2,0) -- (3,2,0) -- (-3,2,0) -- cycle;

        \draw[->,thick,blue] (-2,0,0) -- (-0.3,0,0); \node at (-0.7,0.4,0) {$e^+$};
        \draw[->,thick,blue] (2,0,0) -- (0.3,0,0); \node at (0.7,-0.4,0) {$e^-$};
        
        \draw[->,thick,blue] (1.5/5,1/5,0) -- (1.5,1,0); \node at (1.6,1.2,-0.3) {$\mathcal{B}$};
        \draw[->,thick,blue] (-1.5/5,-1/5,0) -- (-1.5,-1,0); \node at (-1.3,-0.8,-0.3) {$\bar{\mathcal{B}}$};
        
        \draw[thick] (0.8,0,0) arc (0:30:0.8); 
        \node at (1.2,0.3,0) {$\theta_1$};
        
        \draw[->,thick,red] (1.5,1,0) -- (1.5-1/3,1+1.5/3,0) node[left] {$\hat{x}_1$};
        \draw[->,thick,red] (1.5,1,0) -- (1.5,1,0+3.601/6) node[above] {$\hat{y}_1$};
        \draw[->,thick,red] (1.5,1,0) -- (1.5+1.5/3,1+1/3,0) node[right] {$\hat{z}_1$};
        
        \draw[->,thick,red] (-1.5,-1,0) -- (-1.5-1/3,-1+1.5/3,0) node[left] {$\hat{x}_2$};
        \draw[->,thick,red] (-1.5,-1,0) -- (-1.5,-1,0-3.601/6) node[below] {$\hat{y}_2$};
        \draw[->,thick,red] (-1.5,-1,0) -- (-1.5-1.5/3,-1-1/3,0) node[left] {$\hat{z}_2$};

    \end{scope}
\end{tikzpicture}
    \caption{Orientation of the axes in baryon \(\mathcal{B}\) and antibaryon \(\bar{\mathcal{B}}\) helicity frame.}
    \label{fig:scattering_plane}
\end{figure}
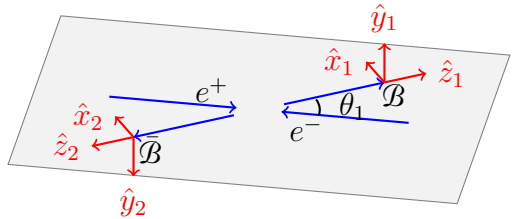

In each helicity frame, the \(\hat{z}_1\) (\(\hat{z}_2\)) axis is aligned along the direction of motion of the \(\mathcal{B}\) (\(\bar{\mathcal{B}}\)) baryon in the CM frame. As illustrated in Fig. \ref{fig:scattering_plane}, the \(\hat{y}_1\) (\(\hat{y}_2\)) axis is perpendicular to the production plane, while the \(\hat{x}_1\) (\(\hat{x}_2\)) axis is chosen to satisfy the right-hand rule.

The initial density matrix of two spin-1/2 baryon-antibaryon pair from the \(e^+e^-\to  \mathcal{B}\bar{\mathcal{B}}\) process can be expressed as\cite{Tabakin:1985yv}:

\begin{equation}
\rho_{\mathcal{B}, \bar{\mathcal{B}}} = \frac{1}{4} \sum_{\mu, \nu = 0}^3 C_{\mu \nu} (\theta_1;\alpha_\psi,\Delta\Phi) \sigma_{\mu}^{\mathcal{B}} \otimes \sigma_{\nu}^{\bar{\mathcal{B}}},
\label{eq:density matrix}
\end{equation}
where \(\sigma_{0}^{\mathcal{B}} = \mathbb{I}_2\), \(\sigma_{1}^{\mathcal{B}} = \sigma_{x}\), \(\sigma_{2}^{\mathcal{B}} = \sigma_{y}\), and \(\sigma_{3}^{\mathcal{B}} = \sigma_{z}\) represent the identity and Pauli matrices. The parameter \( \theta_1 \) represents the scattering angle, while \( \alpha_\psi \) and \( \Delta\Phi \) are two real parameters associated with two production amplitudes. Specifically, the parameter \( \alpha_\psi \) characterizes the angular distribution of the baryon-antibaryon pair production, whereas \( \Delta\Phi \) denotes the relative phase between the two production amplitudes\cite{Perotti:2018wxm, BESIII:2021ypr}. 
In the process \(e^+e^-\to J/\psi,\psi' \to \mathcal{B}\bar{\mathcal{B}}\) where \(\mathcal{B}\) and \(\bar{\mathcal{B}}\) are spin-1/2 baryon and antibaryon, the density matrix \(\rho_{\mathcal{B} \bar{\mathcal{B}}}\) of \(\mathcal{B}\) and \(\bar{\mathcal{B}}\) in their own helicity frame can be given by \eqref{eq:density matrix} and only eight coefficients \(C_{\mu\nu}\) are nonzero. Regardless of the normalization and the overall phase, they are given by\cite{Perotti:2018wxm}

\begin{align*}
C_{00}&=2(1+\alpha_\psi \cos^2\theta_1), \\
C_{02}&=2\sqrt{1-\alpha_\psi^2}\sin{\theta_1}\cos\theta_1\sin(\Delta\Phi), \\
C_{11}&=2\sin^2\theta_1,\\
C_{13}&=2\sqrt{1-\alpha_\psi^2}\sin\theta_1\cos\theta_1\cos(\Delta\Phi),\\
C_{20}&=-C_{02},\\
C_{22}&=\alpha_\psi C_{11},\\
C_{31}&=-C_{13},\\
C_{33}&=-2(\alpha_\psi +\cos^2\theta_1).
\end{align*}
So we can obtain the initial density matrix of \(\mathcal{B}\bar{\mathcal{B}}\) with scarttering angle \(\theta_1\) and two parameters \(\alpha_\psi\), \(\Delta\Phi\) from the experiment data.

Then we consider the processes in which the spin-1/2 \(\mathcal{B}\) decays to a spin-0 meson and a spin-1/2 baryon \(\mathcal{b}\), and the spin-1/2 \(\bar{\mathcal{B}}\) also decays to a spin-0 meson and a spin-1/2 antibaryon \(\bar{\mathcal{b}}\). There is no information of the two spin-0 mesons in the spin space of the final state, so we only need to consider the helicity state of the spin-1/2 baryon and antibaryon daughter particles. This decay process can be described by introducing a \(4\times4\) matrix, and the full expression can be found in \(a_{\mu\nu}\)\cite{Perotti:2018wxm}:
\begin{equation}
    \sigma_\mu^\mathcal{B} \to \sum_{\nu=0}^3 a_{\mu\nu}^\mathcal{B}\sigma_\nu^\mathcal{b},
\end{equation}
in which the \(\sigma^\mathcal{B}_\mu\) matrices are in the mother helicity frame and \(\sigma_\nu^\mathcal{b}\) in the daughter helicity frame. The elements of the matrix \(a_{\mu\nu}\) are parameterized by the decay parameters \(\alpha_D\) and \(\phi_D\) and depend on the decay angles, i.e. \(a_{\mu\nu}^\mathcal{B} (\theta_\mathcal{b},\phi_\mathcal{b};\alpha_D^\mathcal{B},\phi_D^\mathcal{B})\). The decay parameters \(\alpha_D\) and \(\phi_D\) are defined in \cite{Lee:1957qs}. The decay parameter \(\phi_D\), constrained within \([-\pi,\pi]\), is related to the rotation of the spin vector between mother baryon \(\mathcal{B}\) and daughter baryon \(\mathcal{b}\). The decay parameter \(\alpha_D\), constrained within \([-1,1]\), is determined from the angular distribution asymmetry of the \(\mathcal{B}\) baryon in the \(\mathcal{B}\) baryon rest frame~\cite{Batozskaya:2025ula}.
 Therefore, the density matrix of daughter baryon and antibaryon would read:
\begin{equation}
    \rho_{\mathcal{b},\bar{\mathcal{b}}}=\frac{1}{4} \sum_{\mu, \nu = 0 }^3 \sum_{\alpha,\beta=0}^3 C_{\mu \nu}(\theta_1) a_{\mu\alpha}^{\mathcal{B}} a_{\nu\beta}^{\bar{\mathcal{B}}}  \sigma_{\alpha}^{\mathcal{b}} \otimes \sigma_{\beta}^{\bar{\mathcal{b}}},
\end{equation}

To quantify the potential entanglement enhancement through baryon decays, we will compare the entanglement measures of the initial $\mathcal{B}\bar{\mathcal{B}}$ system $\rho_{\mathcal{B}\bar{\mathcal{B}}}$ and the final $\mathcal{b}\bar{\mathcal{b}}$ system $\rho_{\mathcal{b}\bar{\mathcal{b}}}$. Specifically, we calculate and contrast \textit{Negativity} and \textit{Concurrence} of $\rho_{\mathcal{B}\bar{\mathcal{B}}}$ and $\rho_{\mathcal{b}\bar{\mathcal{b}}}$.

The comparison between $\mathcal{N}(\rho_{\mathcal{B}\bar{\mathcal{B}}})$ and $\mathcal{N}(\rho_{\mathcal{b}\bar{\mathcal{b}}})$ (and similarly for $\mathscr{C}$) will demonstrate whether the decay process can \textit{increase} quantum entanglement -- a phenomenon refered to as \textit{entanglement autodistillation}.

\section{Results}\label{sec:results}
Following the methodology outlined in Section~\ref{sec:process}, we now present the comparison between $\mathcal{N}(\rho_{\mathcal{B}\bar{\mathcal{B}}})$ and $\mathcal{N}(\rho_{\mathcal{b}\bar{\mathcal{b}}})$(and similarly for $\mathscr{C}$). We provide an analytical result from a hypothetical example to demonstrate entanglement autodistillation in Section~\ref{Analytic result}. In Section~\ref{Baryon and antibaryon decay}, we numerically simulate several specific physical processes to observe this phenomenon. In these simulations, the decay angles \((\theta_\mathcal{b},\phi_\mathcal{b};\theta_{\bar{\mathcal{b}}},\phi_{\bar{\mathcal{b}}})\) of baryon and antibaryon in their helicity frame have been adjusted to find the most appropriate final density matrix for observing the phenomenon where entanglement increases after decay.

\subsection{Analytic result of a hypothetical example}\label{Analytic result}
We will show how the state after decay can be more entangled than before with a special hypothetical example. For simplicity, the real parameters \(\alpha_\psi\) and \(\Delta\Phi\) describing the initial baryon-antibaryon state are assumed to be \(0\) and \(\pi/2\), the corresponding initial density matrix is
\[
\rho_{\mathcal{B},\bar{\mathcal{B}}} =
\frac{1}{8}
\begin{bmatrix}
2 - 2\cos^2\theta_1 & -\frac{2i}{\cos\theta_1 \sin\theta_1} & \frac{2i}{\cos\theta_1 \sin\theta_1} & 2\sin^2\theta_1 \\
\frac{2i}{\cos\theta_1 \sin\theta_1} & 2 + 2\cos^2\theta_1 & 2\sin^2\theta_1 & \frac{2i}{\cos\theta_1 \sin\theta_1} \\
-\frac{2i}{\cos\theta_1 \sin\theta_1} & 2\sin^2\theta_1 & 2 + 2\cos^2\theta_1 & -\frac{2i}{\cos\theta_1 \sin\theta_1} \\
2\sin^2\theta_1 & -\frac{2i}{\cos\theta_1 \sin\theta_1} & \frac{2i}{\cos\theta_1 \sin\theta_1} & 2 - 2\cos^2\theta_1
\end{bmatrix}.
\]

The concurrence of the initial state is
\[
\mathscr{C}(\rho_{\mathcal{B},\bar{\mathcal{B}}})=\frac{1}{2}(1-|\cos{(2\theta_1)}|)
\]
The decay parameters \(\alpha_D^\mathcal{B}\) and \(\phi_D^\mathcal{B}\) of baryon decay process are assumed to be \(1/2\) and \(\pi/2\). We assume that there is no charge-parity symmetry breaking in this process. Therefore, the antibaryon decay parameters \(\alpha_D^{\bar{\mathcal{B}}}\) and \(\phi_D^{\bar{\mathcal{B}}}\) are assumed to be \(-1/2\) and \(\pi/2\). We choose \(\phi_\mathcal{b}=\phi_{\bar{\mathcal{b}}}=\pi/2\), which denotes that the plane of the final baryon-antibaryon pair is perpendicular to the scattering plane of \(e^+e^- \to \mathcal{B}\bar{\mathcal{B}}\) process. Consider the situation where the scattering angle is \(\theta_1=\pi/4\), and the concurrence of the initial state is \(1/4\). In this situation, the full expression of the concurrence of the final baryon-antibaryon pair \(\mathcal{b}\bar{\mathcal{b}}\), which, depending on the decay angles \(\theta_\mathcal{b}\) and \(\theta_{\bar{\mathcal{b}}}\), is

\[\mathscr{C}(\rho_{\mathcal{b},\bar{\mathcal{b}}})= 3 \sqrt{\frac{1}{(\cos\theta_\mathcal{b} \cos\theta_{\bar{\mathcal{b}}} - 2 (-4 + \sin\theta_\mathcal{b} + \sin\theta_{\bar{\mathcal{b}}}))^2}}. \]
The concurrence of the final state can reach its maximum value \(3/4\) at \(\theta_\mathcal{b}=\theta_{\bar{\mathcal{b}}}=\pi/2\), which is greater than the initial state concurrence \(1/4\).

\subsection{Baryon and antibaryon decay}\label{Baryon and antibaryon decay}
We will consider different specific processes to observe the phenomenon where entanglement increases after decay. Specifically, we will analyze the processes that \(e^+e^-\to J/\psi \to \mathcal{B}\bar{\mathcal{B}}\) and \(e^+e^-\to \psi(3686) \to \mathcal{B}\bar{\mathcal{B}}\), then as we mentioned before, the spin-1/2 \(\mathcal{B}\) decays to a spin-0 meson and a spin-1/2 baryon, and the spin-1/2 \(\bar{\mathcal{B}}\) also decay to a spin-0 meson and a spin-1/2 antibaryon.

\subsubsection{\texorpdfstring{\(J/\psi \to \Xi^- + \bar{\Xi}^+\)}{J/psi -> Xi- + anti-Xi+} then \texorpdfstring{\(\Xi^-\to \Lambda \pi^-\)}{Xi- -> Lambda pi-} and \texorpdfstring{\(\bar{\Xi}^+ \to \bar{\Lambda} \pi^+\)}{nati-Xi+ -> anti-Lambda pi+}}\label{case1}

The two parameters defining the helicity amplitudes of the process that \(e^+e^-\to J/\psi \to \Xi^- \bar{\Xi}^+\)\cite{BESIII:2021ypr}:
\begin{equation}
    \alpha = 0.586 \pm 0.012|_{\text{stat}} \pm 0.010|_{\text{syst}} \quad \text{and} \quad \Delta \Phi = 1.213 \pm 0.046|_{\text{stat}} \pm 0.016|_{\text{syst}}.
\end{equation}

The decay parameters \(\alpha_D\) and \(\phi_D\) of \(\Xi^-\to \Lambda \pi^-\) are given by\cite{ParticleDataGroup:2024cfk}:
\begin{equation}
    \alpha_D=-0.390\pm0.007 \quad \text{and} \quad \phi_D=-1.2\pm1.0^\circ.
\end{equation}

Similarly, the decay parameters \(\bar{\alpha}_D\) and \(\bar{\phi}_D\) of \(\bar{\Xi}^+ \to \bar{\Lambda} \pi^+\) are given by\cite{ParticleDataGroup:2024cfk}:
\begin{equation}
    \bar{\alpha}_D=0.371\pm0.007 \quad \text{and} \quad \bar{\phi}_D=-1.2\pm1.2^\circ.
\end{equation}

\begin{figure}[H]
    \centering
    \begin{minipage}{0.45\textwidth}
        \centering
        \includegraphics[width=\textwidth]{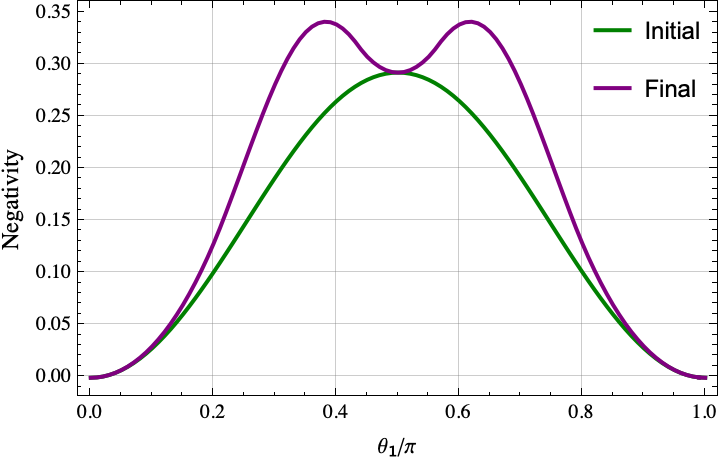}
        \subcaption{Negativity of \(\Xi^-\bar{\Xi}^+\) and \(\Lambda\bar{\Lambda}\)} 
    \end{minipage}\hfill
    \begin{minipage}{0.45\textwidth}
        \centering
        \includegraphics[width=\textwidth]{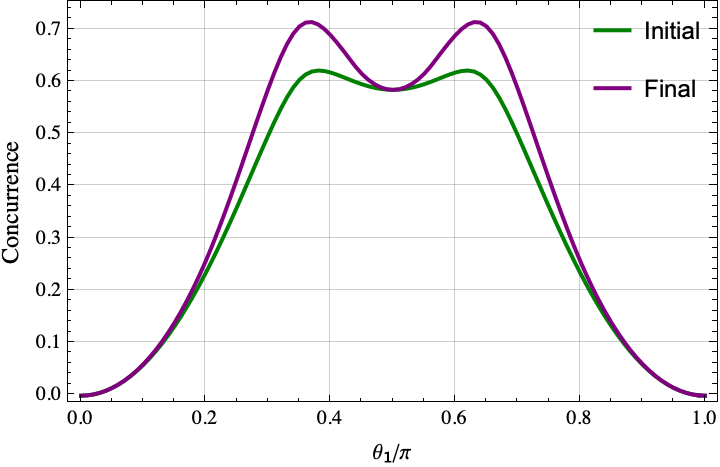}
        \subcaption{Concurrence of \(\Xi^-\bar{\Xi}^+\) and \(\Lambda\bar{\Lambda}\)}
    \end{minipage}

    \caption{
    (a) Negativity of \(\Xi^-\bar{\Xi}^+\) from \(e^+e^- \to J/\psi \to \Xi^- + \bar{\Xi}^+\) process and Negativity of \(\Lambda\bar{\Lambda}\) after \(\Xi^-\to \Lambda \pi^-\) and \(\bar{\Xi}^+ \to \bar{\Lambda} \pi^+\) with the most appropriate decay angle to maximize negativity.
    (b) Concurrence of \(\Xi^-\bar{\Xi}^+\) from \(e^+e^- \to J/\psi \to \Xi^- + \bar{\Xi}^+\) process and Concurrence of \(\Lambda\bar{\Lambda}\) after \(\Xi^-\to \Lambda \pi^-\) and \(\bar{\Xi}^+ \to \bar{\Lambda} \pi^+\) with the most appropriate decay angle to maximize concurrence.
    }
    \label{fig:ent5.1}
\end{figure}

As shown in Fig. \ref{fig:ent5.1}, negativity and concurrence depend on the scattering angle \(\theta_1\) of the \(e^+e^-\to \Xi^-\bar{\Xi}^+\) process. Our calculation of the concurrence of the \(\Xi^-\bar{\Xi}^+\) state is consistent with the result reported in Ref.~\cite{Fabbrichesi:2024rec}. We note that the helicity state of \(\Xi^-\bar{\Xi}^+\) is a separable mixed state (i.e. \(\rho=\sum_i p_i \rho_i^{(1)}\otimes\rho_i^{(2)}\) and there is no entanglement) when \(\theta_1=0 \text{ or }\pi\). The final state of \(\Lambda\bar{\Lambda}\) is also a separable mixed state if there is no entanglement of \(\Xi^-\bar{\Xi}^+\). When \(\theta_1=\pi/2\), although the concurrence and negativity of \(\rho_{\Xi^-\bar{\Xi}^+}\) almost reach the maximum, they do not increase at this point. This phenomenon is much more apparent in Fig. \ref{fig:incent5.1} as shown below.

\begin{figure}[H]
    \centering
    \begin{minipage}{0.45\textwidth}
        \centering
        \includegraphics[width=\textwidth]{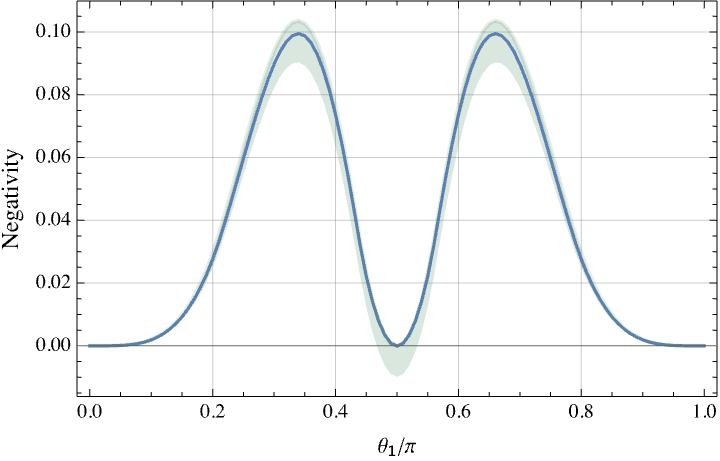}
        \subcaption{Increase of negativity} 
    \end{minipage}\hfill
    \begin{minipage}{0.45\textwidth}
        \centering
        \includegraphics[width=\textwidth]{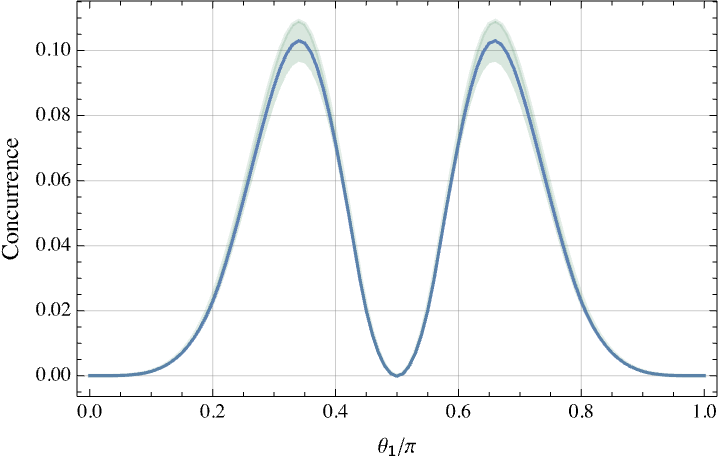}
        \subcaption{Increase of concurrence}
    \end{minipage}

    \caption{(a) Increase of negativity after \(\Xi^-\to \Lambda \pi^-\) and \(\bar{\Xi}^+ \to \bar{\Lambda} \pi^+\) with the most appropriate decay angle to maximize negativity. (b) Increase of concurrence after \(\Xi^-\to \Lambda \pi^-\) and \(\bar{\Xi}^+ \to \bar{\Lambda} \pi^+\) with the most appropriate decay angle to maximize concurrence.}
    \label{fig:incent5.1}
\end{figure}

The total uncertainty in our results arises from two primary sources. The first is the uncertainty in the experimental input parameters. The second is the statistical error associated with our numerical method, which is based on uniform random sampling; using a finite number of sampling points may lead to fluctuations in the results. The gray band in Fig.~\ref{fig:incent5.1} illustrates the first type of uncertainty, which originates from the uncertainties in the input parameters \(\alpha\) and \(\Delta\Phi\) that determine the initial density matrix, as well as from the uncertainties in the decay parameters \(\alpha_D\) and \(\bar{\alpha}_D\). This uncertainty does not depend on the decay parameters \(\phi_D\) and \(\bar{\phi}_D\), as will be explained in Section~\ref{sec:mechanism}.

\subsubsection{\texorpdfstring{\(\psi(3686) \to \Xi^- + \bar{\Xi}^+\)}{psi(3686) -> Xi- + anti-Xi+} then \texorpdfstring{\(\Xi^-\to \Lambda \pi^-\)}{Xi- -> Lambda pi-} and \texorpdfstring{\(\bar{\Xi}^+ \to \bar{\Lambda} \pi^+\)}{nati-Xi+ -> anti-Lambda pi+}}\label{case2}

The two parameters defining the helicity amplitudes of the process that \(e^+e^-\to \psi(3686) \to \Xi^-  \bar{\Xi}^+\)\cite{BESIII:2022lsz}:
\begin{equation}
    \alpha = 0.693 \pm 0.048|_{\text{stat}} \pm 0.049|_{\text{syst}} \quad \text{and} \quad \Delta \Phi = 0.667 \pm 0.111|_{\text{stat}} \pm 0.058|_{\text{syst}}.
\end{equation}

\begin{figure}[H]
    \centering
    \begin{minipage}{0.45\textwidth}
        \centering
        \includegraphics[width=\textwidth]{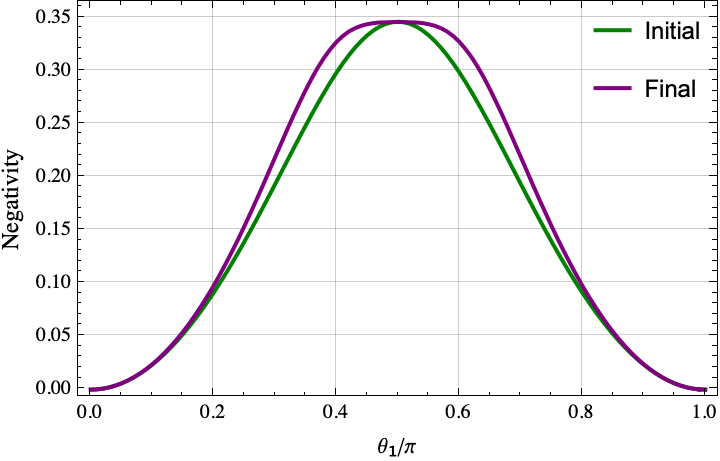}
        \subcaption{Negativity of \(\Xi^-\bar{\Xi}^+\) and \(\Lambda\bar{\Lambda}\)} 
    \end{minipage}\hfill
    \begin{minipage}{0.45\textwidth}
        \centering
        \includegraphics[width=\textwidth]{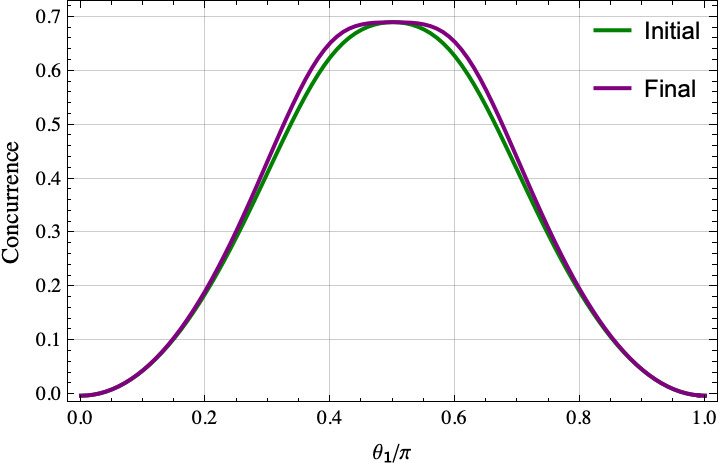}
        \subcaption{Concurrence of \(\Xi^-\bar{\Xi}^+\) and \(\Lambda\bar{\Lambda}\)}
    \end{minipage}

    \caption{
    (a) Negativity of \(\Xi^-\bar{\Xi}^+\) from \(e^+e^- \to \psi(3686) \to \Xi^- + \bar{\Xi}^+\) process and Negativity of \(\Lambda\bar{\Lambda}\) after \(\Xi^-\to \Lambda \pi^-\) and \(\bar{\Xi}^+ \to \bar{\Lambda} \pi^+\) with the most appropriate decay angle to maximize negativity.
    (b) Concurrence of \(\Xi^-\bar{\Xi}^+\) from \(e^+e^- \to \psi(3686) \to \Xi^- + \bar{\Xi}^+\) process and Concurrence of \(\Lambda\bar{\Lambda}\) after \(\Xi^-\to \Lambda \pi^-\) and \(\bar{\Xi}^+ \to \bar{\Lambda} \pi^+\) with the most appropriate decay angle to maximize concurrence.
    }
    \label{fig:ent5.2}
\end{figure}

\begin{figure}[H]
    \centering
    \begin{minipage}{0.45\textwidth}
        \centering
        \includegraphics[width=\textwidth]{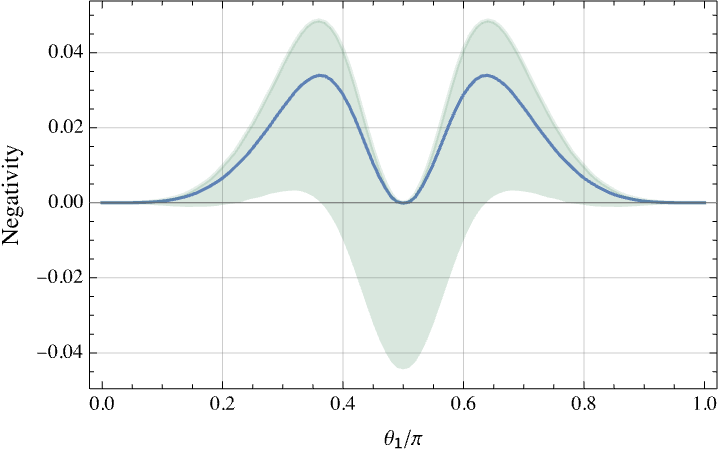}
        \subcaption{Increase of negativity} 
    \end{minipage}\hfill
    \begin{minipage}{0.45\textwidth}
        \centering
        \includegraphics[width=\textwidth]{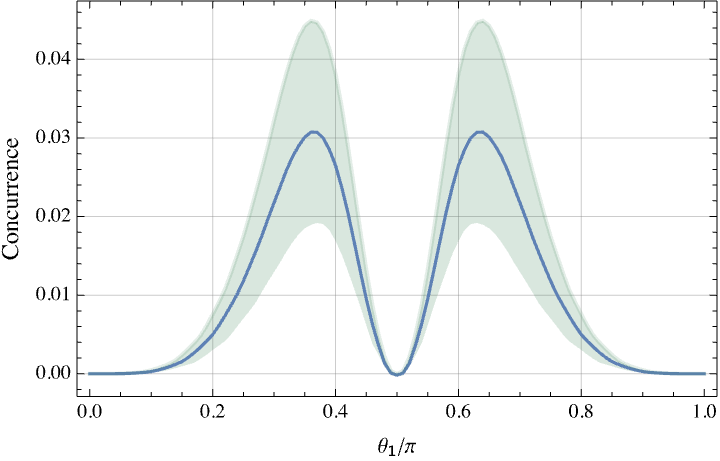}
        \subcaption{Increase of concurrence}
    \end{minipage}

    \caption{(a) Increase of negativity after \(\Xi^-\to \Lambda \pi^-\) and \(\bar{\Xi}^+ \to \bar{\Lambda} \pi^+\) with the most appropriate decay angle to maximize negativity. (b) Increase of concurrence after \(\Xi^-\to \Lambda \pi^-\) and \(\bar{\Xi}^+ \to \bar{\Lambda} \pi^+\) with the most appropriate decay angle to maximize concurrence.}
    \label{fig:incent5.2}
\end{figure}

As shown in Fig.~\ref{fig:ent5.2}, the entanglement behavior of the process \(\psi(3686) \to \Xi^- + \bar{\Xi}^+\), followed by the decays \(\Xi^- \to \Lambda \pi^-\) and \(\bar{\Xi}^+ \to \bar{\Lambda} \pi^+\), exhibits a pattern similar to that seen in Fig.~\ref{fig:ent5.1}. The negativity and concurrence of the \(\Xi^-\bar{\Xi}^+\) state depend on the scattering angle \(\theta_1\), and the entanglement vanishes at \(\theta_1 = 0\) or \(\pi\), where the helicity state becomes separable. Our results for the initial-state concurrence are consistent with those reported in Ref.~\cite{Fabbrichesi:2024rec}. Compared to the \(J/\psi\) channel in Fig.~\ref{fig:ent5.1}, the increase of entanglement after decay is less pronounced, as can be seen in Fig.~\ref{fig:incent5.2}. Additionally, Fig.~\ref{fig:incent5.2} exhibits larger uncertainties in the post-decay entanglement, mainly due to the uncertainties in the input parameters \(\alpha\) and \(\Delta\Phi\).

\subsubsection{\texorpdfstring{\(J/\psi \to \Xi^0 + \bar{\Xi}^0\)}{J/psi -> Xi0 + anti-Xi0} then \texorpdfstring{\(\Xi^0\to \Lambda \pi^0\)}{Xi0 -> Lambda pi0} and \texorpdfstring{\(\bar{\Xi}^0 \to \bar{\Lambda} \pi^0\)}{anti-Xi0 -> anti-Lambda pi0}}\label{case3}

The two parameters defining the helicity amplitudes of the process that \(e^+e^-\to J/\psi \to \Xi^0 \bar{\Xi}^0\)\cite{BESIII:2023drj}:
\begin{equation}
    \alpha = 0.514 \pm 0.006|_{\text{stat}} \pm 0.0015|_{\text{syst}} \quad \text{and} \quad \Delta \Phi = 1.168 \pm 0.019|_{\text{stat}} \pm 0.018|_{\text{syst}}.
\end{equation}

The decay parameters \(\alpha_D\) and \(\phi_D\) of \(\Xi^0\to \Lambda \pi^0\) are given by\cite{ParticleDataGroup:2024cfk}:
\begin{equation}
    \alpha_D=-0.349\pm0.009 \quad \text{and} \quad \phi_D=0.3\pm0.6^\circ.
\end{equation}

Similarly, the decay parameters \(\bar{\alpha}_D\) and \(\bar{\phi}_D\) of \(\bar{\Xi}^0 \to \bar{\Lambda} \pi^0\) are given by\cite{ParticleDataGroup:2024cfk}:
\begin{equation}
    \bar{\alpha}_D=0.379\pm0.004 \quad \text{and} \quad \bar{\phi}_D=-0.3\pm0.6^\circ.
\end{equation}

\begin{figure}[H]
    \centering
    \begin{minipage}{0.45\textwidth}
        \centering
        \includegraphics[width=\textwidth]{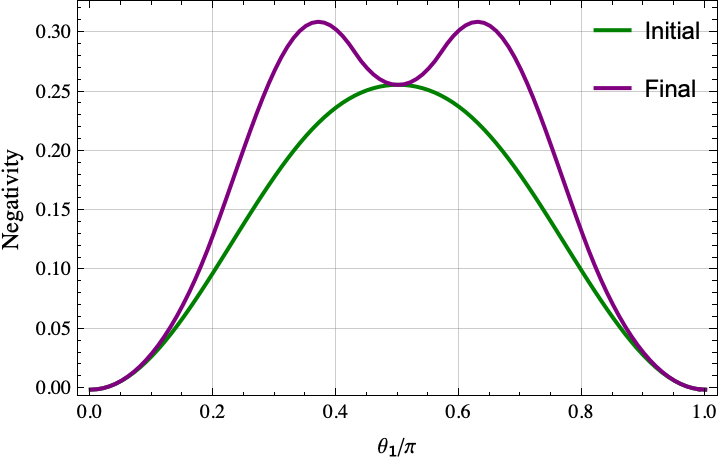}
        \subcaption{Negativity of \(\Xi^0\bar{\Xi}^0\) and \(\Lambda\bar{\Lambda}\)} 
    \end{minipage}\hfill
    \begin{minipage}{0.45\textwidth}
        \centering
        \includegraphics[width=\textwidth]{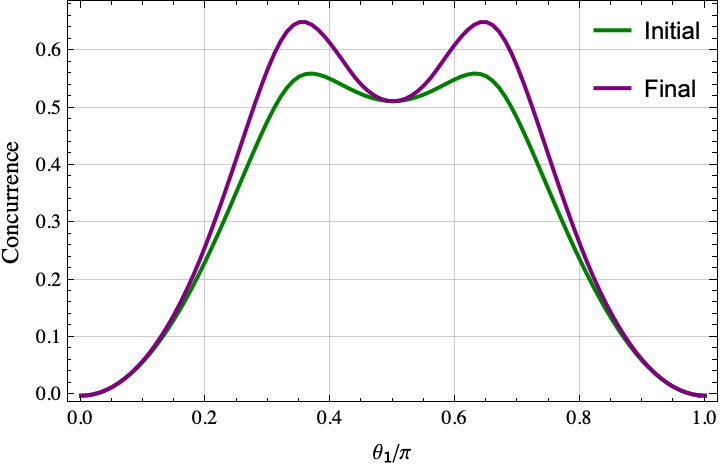}
        \subcaption{Concurrence of \(\Xi^0\bar{\Xi}^0\) and \(\Lambda\bar{\Lambda}\)}
    \end{minipage}

    \caption{
    (a) Negativity of \(\Xi^0\bar{\Xi}^0\) from \(e^+e^- \to J/\psi \to \Xi^0 + \bar{\Xi}^0\) process and Negativity of \(\Lambda\bar{\Lambda}\) after \(\Xi^0\to \Lambda \pi^0\) and \(\bar{\Xi}^0 \to \bar{\Lambda} \pi^0\) with the most appropriate decay angle to maximize negativity.
    (b) Concurrence of \(\Xi^0\bar{\Xi}^0\) from \(e^+e^- \to J/\psi \to \Xi^0 + \bar{\Xi}^0\) process and Concurrence of \(\Lambda\bar{\Lambda}\) after \(\Xi^0\to \Lambda \pi^0\) and \(\bar{\Xi}^0 \to \bar{\Lambda} \pi^0\) with the most appropriate decay angle to maximize concurrence.
    }
    \label{fig:ent4.1}
\end{figure}

\begin{figure}[H]
    \centering
    \begin{minipage}{0.45\textwidth}
        \centering
        \includegraphics[width=\textwidth]{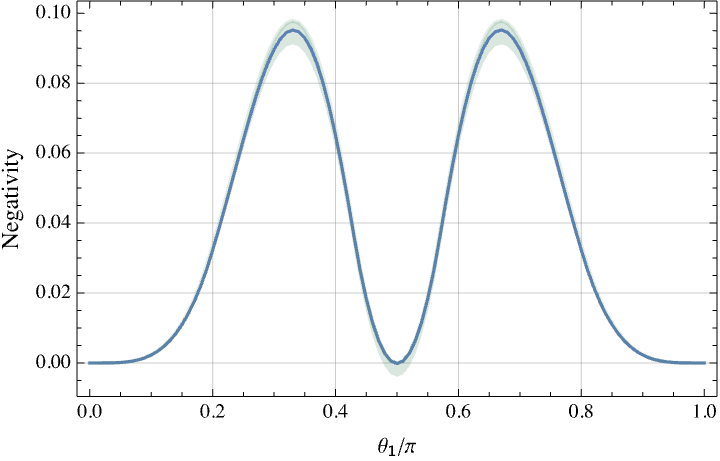}
        \subcaption{Increase of negativity} 
    \end{minipage}\hfill
    \begin{minipage}{0.45\textwidth}
        \centering
        \includegraphics[width=\textwidth]{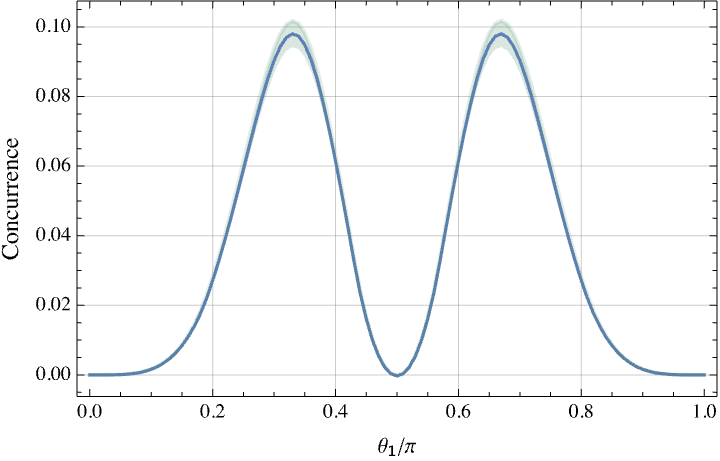}
        \subcaption{Increase of concurrence}
    \end{minipage}

    \caption{(a) Increase of negativity after \(\Xi^0\to \Lambda \pi^0\) and \(\bar{\Xi}^0 \to \bar{\Lambda} \pi^0\) with the most appropriate decay angle to maximize negativity. (b) Increase of concurrence after \(\Xi^0\to \Lambda \pi^0\) and \(\bar{\Xi}^0 \to \bar{\Lambda} \pi^0\) with the most appropriate decay angle to maximize concurrence.}
    \label{fig:incent4.1}
\end{figure}

As shown in Fig.~\ref{fig:ent4.1}, both the negativity and concurrence of the \(\Xi^0\bar{\Xi}^0\) state produced in the process \(e^+e^- \to J/\psi \to \Xi^0 + \bar{\Xi}^0\) also exhibit dependence on the scattering angle \(\theta_1\). Similar to the charged case, our results are consistent with those reported in Ref.~\cite{Fabbrichesi:2024rec}. In particular, the entanglement vanishes when \(\theta_1 = 0\) or \(\pi\), corresponding to a separable helicity state. Consequently, the final state of \(\Lambda\bar{\Lambda}\), produced via \(\Xi^0 \to \Lambda \pi^0\) and \(\bar{\Xi}^0 \to \bar{\Lambda} \pi^0\), remains unentangled under these conditions. At \(\theta_1 = \pi/2\), the entanglement of the \(\Xi^0\bar{\Xi}^0\) state nearly reaches its maximum; however, the entanglement of the final state does not increase at this point, a feature that is clearly illustrated in Fig.~\ref{fig:incent4.1}.

As discussed earlier, the decay angles \((\theta_\mathcal{b},\phi_\mathcal{b};\theta_{\bar{\mathcal{b}}},\phi_{\bar{\mathcal{b}}})\) of the baryon and antibaryon in their respective helicity frames have been adjusted to identify the optimal final density matrix that reveals the entanglement amplification phenomenon. In most cases, multiple sets of decay angles can lead to the maximum value of entanglement after the decay process. However, in the scenarios shown in Sections~\ref{case1} and~\ref{case3}, there exist unique decay angles corresponding to specific scattering angle intervals, as revealed by numerical simulations. Specifically, when the scattering angle \(\theta_1 \in [0.22\pi, 0.42\pi]\), the unique configuration is  
\[
(\theta_\mathcal{b},\phi_\mathcal{b};\theta_{\bar{\mathcal{b}}},\phi_{\bar{\mathcal{b}}}) = \left(\tfrac{1}{2}\pi,\tfrac{3}{2}\pi;\tfrac{1}{2}\pi,\tfrac{3}{2}\pi\right),
\]
and when \(\theta_1 \in [0.58\pi, 0.78\pi]\), the unique configuration becomes  
\[
(\theta_\mathcal{b},\phi_\mathcal{b};\theta_{\bar{\mathcal{b}}},\phi_{\bar{\mathcal{b}}}) = \left(\tfrac{1}{2}\pi,\tfrac{1}{2}\pi;\tfrac{1}{2}\pi,\tfrac{1}{2}\pi\right).
\]
Fortunately, as illustrated in Fig.~\ref{fig:incent5.1} and Fig.~\ref{fig:incent4.1}, these regions coincide with the optimal conditions for observing entanglement enhancement after decay.

\section{Reconstruction of Spin Configuration}\label{sec:reconstruction}
In this section, we present a method to reconstruct the spin configuration of the mother particles by analyzing the angular distributions of their decay products. This reconstruction is crucial for experimentally probing entanglement autodistillation via measurable momentum correlations. As a concrete example, we consider the decay chain \(J/\psi \to \Xi^- + \bar{\Xi}^+\), followed by \(\Xi^- \to \Lambda \pi^-\) and \(\bar{\Xi}^+ \to \bar{\Lambda} \pi^+\), and finally \(\Lambda \to p \pi^-\) and \(\bar{\Lambda} \to \bar{p} \pi^+\). The spin configuration of the \(\Xi^- \bar{\Xi}^+\) system can be reconstructed from the angular distributions of the \(\Lambda \bar{\Lambda}\) pair, and in turn, the spin configuration of the \(\Lambda \bar{\Lambda}\) system can be inferred from the angular distributions of the \(p \bar{p}\) pair. This hierarchical structure of decays allows one to track and reconstruct spin information step by step, providing an experimentally accessible pathway to test entanglement features in the hadronic system.

From Section~\ref{sec:process}, we know that the spin configuration of the \(\Xi^- \bar{\Xi}^+\) system can be described by a density matrix:
\[
\rho_{\Xi^-, \bar{\Xi}^+} = \frac{1}{4} \sum_{\mu, \nu = 0}^3 C_{\mu \nu}  \sigma_{\mu}^{\Xi^-} \otimes \sigma_{\nu}^{\bar{\Xi}^+},
\]
where the coefficients \(C_{\mu\nu}\) fully characterize the spin correlation. Therefore, reconstructing the spin configuration reduces to determining the values of \(C_{\mu\nu}\).

We now consider the angular distribution of the \(\Lambda \bar{\Lambda}\) system. Its unnormalized density matrix is given by
\begin{equation}
    \rho_{\Lambda,\bar{\Lambda}}=\frac{1}{4} \sum_{\mu, \nu = 0 }^3 \sum_{\alpha,\beta=0}^3 C_{\mu \nu} a_{\mu\alpha}^{\Xi^-} a_{\nu\beta}^{\bar{\Xi}^+}  \sigma_{\alpha}^{\Lambda} \otimes \sigma_{\beta}^{\bar{\Lambda}}= \frac{1}{4}\sum_{\alpha,\beta=0}^{3}C^{\Lambda\bar{\Lambda}}_{\alpha\beta} \sigma_{\alpha}^{\Lambda} \otimes \sigma_{\beta}^{\bar{\Lambda}},
    \label{Lambda density matrix}
\end{equation}
where the \(C^{\Lambda\bar{\Lambda}}_{\alpha\beta}\) are defined as the coefficients of \(\sigma_{\alpha}^{\Lambda} \otimes \sigma_{\beta}^{\bar{\Lambda}}\) and the \(a_{\mu\nu}\) are defined as in~\cite{Perotti:2018wxm}:
\[
\begin{aligned}
a_{00} &= 1, \\
a_{10} &= \alpha_D \cos \phi \sin \theta, \\
a_{20} &= \alpha_D \sin \theta \sin \phi, \\
a_{30} &= \alpha_D \cos \theta, \\
a_{03} &= \alpha_D, \\
a_{11} &= \gamma_D \cos \theta \cos \phi - \beta_D \sin \phi, \\
a_{12} &= -\beta_D \cos \theta \cos \phi - \gamma_D \sin \phi, \\
a_{13} &= \sin \theta \cos \phi, \\
a_{21} &= \beta_D \cos \phi + \gamma_D \cos \theta \sin \phi, \\
a_{22} &= \gamma_D \cos \phi - \beta_D \cos \theta \sin \phi, \\
a_{23} &= \sin \theta \sin \phi, \\
a_{31} &= -\gamma_D \sin \theta, \\
a_{32} &= \beta_D \sin \theta, \\
a_{33} &= \cos \theta.
\end{aligned}
\]
Where \(\beta_D\) defined as \((1-\alpha_D^2)^{1/2}\sin \phi_D\) and \(\gamma_D\) defined as \((1-\alpha_D^2)^{1/2}\cos \phi_D\)\cite{ParticleDataGroup:2024cfk}.
This unnormalized density matrix contains two types of information. First, the spin configuration of the \(\Lambda \bar{\Lambda}\) system, which can be obtained by normalizing the matrix to remove the effects of the angular distribution. Second, the angular distribution of the \(\Lambda \bar{\Lambda}\) system, which is obtained by tracing out the spin degrees of freedom:
\[
f_{\Lambda\bar{\Lambda}}(\theta_\Lambda,\phi_\Lambda;\theta_{\bar{\Lambda}},\phi_{\bar{\Lambda}})\propto\mathrm{Tr}(\rho_{\Lambda,\bar{\Lambda}})=\sum_{\mu, \nu = 0 }^3 C_{\mu \nu} a_{\mu0}^{\Xi^-}(\theta_\Lambda,\phi_\Lambda) a_{\nu0}^{\bar{\Xi}^+}(\theta_{\bar{\Lambda}},\phi_{\bar{\Lambda}}).
\]
Here, \(f_{\Lambda\bar{\Lambda}}(\theta_\Lambda,\phi_\Lambda;\theta_{\bar{\Lambda}},\phi_{\bar{\Lambda}})\) denotes the angular distribution function in the helicity frame.

We notice \(\{a_{\mu0}\}\) form an orthogonal set in function space.
Therefore, the coefficients \(C_{\mu\nu}\) can be extracted by integrating \(f_{\Lambda\bar{\Lambda}}\) with suitable \(a_{\mu0}\),
\[
C_{\mu\nu} \propto \int f_{\Lambda\bar{\Lambda}}(\theta_\Lambda,\phi_\Lambda;\theta_{\bar{\Lambda}},\phi_{\bar{\Lambda}})a_{\mu0}^{\Xi^-} a_{\nu0}^{\bar{\Xi}^+} \, \mathrm{d}\theta_\Lambda \mathrm{d}\phi_\Lambda \mathrm{d}\theta_{\bar{\Lambda}} \mathrm{d}\phi_{\bar{\Lambda}}.
\]

Similarly, the spin density matrix of the \(p\bar{p}\) system is given by:
\[
\rho_{p\bar{p}}=\frac{1}{4} \sum_{\mu, \nu = 0 }^3 \sum_{\alpha,\beta=0}^3 \sum_{\alpha',\beta'=0}^3 C_{\mu \nu} \, a_{\mu\alpha}^{\Xi^-} a_{\nu\beta}^{\bar{\Xi}^+} a_{\alpha\alpha'}^{\Lambda} a_{\beta\beta'}^{\bar{\Lambda}} \, \sigma_{\alpha'}^{p} \otimes \sigma_{\beta'}^{\bar{p}}.
\]

To determine the coefficients in Eq.~\eqref{Lambda density matrix}, we consider the angular distribution of the final-state \(p\bar{p}\) pair, which is obtained by tracing over the spin degrees of freedom:
\[
f_{p\bar{p}}(\theta_p,\phi_p;\theta_{\bar{p}},\phi_{\bar{p}})\propto \mathrm{Tr}(\rho_{p\bar{p}})=\sum_{\mu, \nu = 0 }^3 \sum_{\alpha,\beta=0}^3 C_{\mu \nu} \, a_{\mu\alpha}^{\Xi^-} a_{\nu\beta}^{\bar{\Xi}^+} a_{\alpha0}^{\Lambda} a_{\beta0}^{\bar{\Lambda}}.
\]

As before, the coefficients in Eq.~\eqref{Lambda density matrix} can then be extracted by integrating the angular distribution function \(f_{p\bar{p}}\) with appropriate auxiliary functions,
\[
C^{\Lambda\bar{\Lambda}}_{\alpha\beta}=\sum_{\mu,\nu=0}^3 C_{\mu\nu} \, a^{\Xi^-}_{\mu\alpha} \, a^{\bar{\Xi}^+}_{\nu\beta} \propto \int f_{p\bar{p}}(\theta_p,\phi_p;\theta_{\bar{p}},\phi_{\bar{p}}) a_{\alpha0}^{\Lambda} a_{\beta0}^{\bar{\Lambda}}  \, \mathrm{d}\theta_p \, \mathrm{d}\phi_p \, \mathrm{d}\theta_{\bar{p}} \, \mathrm{d}\phi_{\bar{p}}.
\]

\section{Mechanism Behind Entanglement Increase in Particle Decays}\label{sec:mechanism}
In this section, we discuss the underlying mechanism responsible for the spontaneous increase in entanglement observed after the decay of certain systems. For the decay of a spin-1/2 mother particle into a spin-1/2 daughter particle and a spin-0 meson, the relationship between the polarization vectors of the mother and daughter particles can be characterized by the decay parameters \(\alpha_D\), \(\beta_D\), and \(\gamma_D\), as given in Ref.~\cite[p.~1001]{ParticleDataGroup:2024cfk}:
\[
\mathbf{P}_D = \frac{(\alpha_D + \mathbf{P}_M \cdot \hat{\mathbf{n}}) \hat{\mathbf{n}} + \beta_D (\mathbf{P}_M \times \hat{\mathbf{n}}) + \gamma_D \hat{\mathbf{n}} \times (\mathbf{P}_M \times \hat{\mathbf{n}})}{1 + \alpha_D\, \mathbf{P}_M \cdot \hat{\mathbf{n}}},
\]
where \(\mathbf{P}_M\) and \(\mathbf{P}_D\) denote the polarization vectors of the mother and daughter particles in their respective rest frames, and \(\hat{\mathbf{n}}\) is the unit vector in the direction of the daughter particle, defined in the rest frame of the mother particle.
This expression was mentioned in Lee’s seminal work~\cite{Lee:1957qs}, although no explicit derivation was provided there. It can be derived using the density matrix formalism applied to the decay process~\cite{Perotti:2018wxm}, which in turn is based on the helicity formalism developed by Jacob and Wick~\cite{Jacob:1959at}.
A detailed derivation of the polarization formula can be found in the \hyperref[app:polarization]{Appendix}.

Consider a mother particle with spin-up polarization, i.e., \(\mathbf{P}_M = \hat{\mathbf{z}}\). After decay, the polarization of the daughter particle in its helicity frame is given by the vector \((\beta_D \sin{\theta},\, \gamma_D \sin{\theta},\, \alpha_D + \cos{\theta})\), where \(\cos{\theta}=\hat{\mathbf{z}} \cdot \hat{\mathbf{n}}\). Similarly, if the mother particle is spin-down (\(\mathbf{P}_M = -\hat{\mathbf{z}}\)), the daughter particle's polarization becomes \((-\beta_D \sin{\theta},\,-\gamma_D \sin{\theta},\, \alpha_D - \cos{\theta})\).

The decay process can be described by a local operator \(D\). For example, in the decay chain discussed in Section~\ref{sec:reconstruction}, we define \(D^{\Xi^-}\) and \(D^{\bar{\Xi}^+}\), leading to the transformation:
\[
\rho_{\Lambda,\bar{\Lambda}}=(D^{\Xi^-}\otimes D^{\bar{\Xi}^+})^{\dagger} \rho_{\Xi^-, \bar{\Xi}^+} D^{\Xi^-}\otimes D^{\bar{\Xi}^+}\equiv \frac{1}{4} \sum_{\mu, \nu = 0 }^3 \sum_{\alpha,\beta=0}^3 C_{\mu \nu} a_{\mu\alpha}^{\Xi^-} a_{\nu\beta}^{\bar{\Xi}^+}  \sigma_{\alpha}^{\Lambda} \otimes \sigma_{\beta}^{\bar{\Lambda}}.
\]

When $\alpha_D = 0$, there is no parity violation, and the two orthogonal polarization states of the mother particle are mapped to two orthogonal states of the daughter particle. In this case, the decay process corresponds to a local unitary (LU) transformation. To see this, consider a general linear operator $D$ acting on the Hilbert space of the mother particle. Suppose the initial polarization states $|\psi_1\rangle$ and $|\psi_2\rangle$ form an orthonormal basis, and are mapped under $D$ to $D|\psi_1\rangle = |\phi_1\rangle$ and $D|\psi_2\rangle = |\phi_2\rangle$, which remain orthonormal when $\alpha_D = 0$. Since the inner products are preserved, i.e.,
\[
\langle \phi_i | \phi_j \rangle = \langle D\psi_i | D\psi_j \rangle = \langle \psi_i | D^\dagger D | \psi_j \rangle = \delta_{ij},
\]
this implies that $D^\dagger D = I$ on the span of $\{|\psi_1\rangle, |\psi_2\rangle\}$, and hence $D$ is unitary on that subspace. Therefore, the decay process in this case corresponds to a local unitary transformation and does not change the entanglement of the bipartite state.

However, if $\alpha_D \neq 0$, the decay process exhibits parity violation, and the polarization vectors associated with the daughter particles become non-orthogonal, meaning that the corresponding operator $D$ no longer maps an orthonormal basis to another orthonormal basis. In this case, $D$ is not unitary, and the decay process may alter the entanglement of the system.

Now we can explain why the decay parameters \(\phi_D\) and \(\bar{\phi}_D\) do not affect the change in entanglement during the decay process, based on the mechanism discussed above.

For a spin-up and a spin-down mother particle, the corresponding daughter polarizations in their helicity frames are given by the vectors  

\[
(\beta_D \sin{\theta},\, \gamma_D \sin{\theta},\, \alpha_D + \cos{\theta}) \quad \text{and} \quad (-\beta_D \sin{\theta},\,-\gamma_D \sin{\theta},\, \alpha_D - \cos{\theta}),
\]
respectively. These vectors correspond to spin states that can be written as  
\[
\left(\cos\frac{\theta_1}{2},\; e^{i\phi_1} \sin\frac{\theta_1}{2} \right) \quad \text{and} \quad
\left(\cos\frac{\theta_2}{2},\; e^{i\phi_2} \sin\frac{\theta_2}{2} \right),
\]  
where the polar and azimuthal angles are given by

\[
\begin{aligned}
    \tan\phi_1 &= \frac{\gamma_D\sin\theta}{\beta_D\sin\theta} = \frac{\cos \phi_D}{\sin \phi_D}, \\
    \tan\phi_2 &= \frac{-\gamma_D\sin\theta}{-\beta_D\sin\theta} = \frac{\cos \phi_D}{\sin \phi_D}, \\
    \cos\theta_1 &= \frac{\alpha_D + \cos\theta}{1+\alpha_D\cos\theta}, \\
    \cos\theta_2 &= \frac{\alpha_D - \cos\theta}{1 - \alpha_D\cos\theta}.
\end{aligned}
\]
We observe that \(\phi_1 = \phi_2 = \phi = \arctan(\cot(\phi_D))\), which means the decay parameter \(\phi_D\) introduces only a phase difference between the two spin states. This effect can be described by a local unitary transformation:
\[
U = \begin{pmatrix}
1 & 0 \\
0 & e^{i\delta\phi}
\end{pmatrix}.
\]
For instance, in the case considered above, modifying the decay parameter \(\phi_D^{\Xi^-}\) results in a transformed density matrix for the \(\Lambda\bar{\Lambda}\) system:
\[
\rho_{\Lambda,\bar{\Lambda}}' = (U \otimes \mathbb{I}_2)^\dagger \rho_{\Lambda,\bar{\Lambda}} (U \otimes \mathbb{I}_2).
\]
Since local unitary operations do not alter the entanglement of a quantum system, we conclude that \(\phi_D\) and \(\bar{\phi}_D\) do not affect the entanglement. Therefore, their uncertainties can be safely neglected in the entanglement analysis.

\section{Conclusion}
Our analysis demonstrates that the decay of a baryon-antibaryon pair, initially described by an entangled mixed state, can result in an increase in entanglement---a phenomenon referred to as \textit{entanglement autodistillation}.  
Unlike conventional LOCC protocols, which cannot increase entanglement, this effect arises naturally through SLOCC.  
While previous studies~\cite{Aguilar-Saavedra:2024fig} have investigated the evolution of entanglement during decay processes, our work differs in that we focus on scenarios where the accompanying mesons---due to their pseudoscalar nature---do not interfere with the entanglement of the system. In contrast, Ref.~\cite{Aguilar-Saavedra:2024fig} considered the process \(t \to Wb\), where the final-state entanglement is affected by additional degrees of freedom.  
We further examine the underlying mechanism behind entanglement autodistillation and show that this phenomenon depends solely on the initial state and the decay parameter \(\alpha_D\), and is independent of the parameter \(\phi_D\).
Importantly, the autodistillation effect we identify could be experimentally tested at an \(e^+e^-\) collider.

\section*{Acknowledgements}

We are grateful to Zhaofeng Kang, Jianwei Cui, and Ruilin Zhu for their valuable discussions and insightful suggestions, which greatly contributed to the development of this work. Special thanks go to Guomengchao Bian for many enjoyable mathematical conversations that helped broaden our perspective during the course of this research. This work is supported by the Natural Science Foundation of China under Grant No. 12375086 and No.12005070, and Hubei Provincial Natural Science Foundation of China under Grant No.2025AFB557.

\appendix
\section*{Appendix}
\addcontentsline{toc}{section}{Appendix}

\subsection*{Derivation of the Polarization Formula}
\label{app:polarization}

The density matrix of a spin-up mother particle can be expressed as
\[
\rho_{M}=|\uparrow\rangle\langle\uparrow|=
\begin{pmatrix}
1 & 0\\
0 & 0
\end{pmatrix}=\frac{1}{2}\sigma_0+\frac{1}{2}\sigma_3.
\]
With the density matrix formalism mentioned in~\cite{Perotti:2018wxm}, the density matrix of the spin of the final particles in their own helicity frame can be expressed as

\[
\begin{aligned}
    \rho_{D}&=\frac{1}{2}\sum_{\mu=0}^3 a_{0\mu}\sigma_\mu+\frac{1}{2}\sum_{\mu=0}^{3}a_{3\mu}\sigma_\mu\\
    &=\frac{1}{2}[(1+\alpha_D \cos{\theta})\sigma_0-\gamma_D \sin{\theta}\sigma_1+\beta_D\sin{\theta}\sigma_2+(\alpha_D+\cos{\theta})\sigma_3].
\end{aligned}
\]
As we mentioned in Section~\ref{sec:reconstruction}, this unnormalized density matrix contains two parts of information, the spin configuration and the angular distribution of the daughter particle. The spin configuration of the daughter particle can be obtained by normalizing the density matrix
\[
\rho_D'=\frac{\rho_D}{Tr(\rho_D)}=\frac{1}{2}(\sigma_0+\vec{r}\cdot\vec{\sigma}),
\]
where 
\[
\vec{r}=\frac{1}{1+\alpha_D\cos{\theta}}(-\gamma_D\sin{\theta},\beta_D\sin{\theta},\alpha_D+\cos{\theta}).
\]

The density matrix \(\rho_D'\) represents a pure state, and the vector \(\vec{r}\) denotes the polarization direction of the daughter particle~\cite[p.~105]{Nielsen:2012yss}. This result is consistent with the expression given by Lee, in which the daughter polarization direction is \((\beta_D \sin{\theta},\ \gamma_D \sin{\theta},\ \alpha_D + \cos{\theta})\), up to a rotation of \(\pi/2\) around the \(\hat{z}\)-axis between the coordinate systems used. This difference arises solely from the choice of coordinate convention and does not affect any physical predictions.

The polarization formula can also be derived by Feynman's formalism~\cite[pp.~223--225]{Commins:1983ns}. With this method, we can derive the transition rate as
\[
R(\hat{\omega}_i,\hat{\omega}_f,\hat{\mathbf{n}})= 1 + \gamma_D\, \hat{\omega}_f \cdot \hat{\omega}_i + (1 - \gamma_D)(\hat{\omega}_f \cdot \hat{\mathbf{n}})(\hat{\omega}_i \cdot \hat{\mathbf{n}})
+ \alpha_D \left( \hat{\omega}_f \cdot \hat{\mathbf{n}} + \hat{\omega}_i \cdot \hat{\mathbf{n}} \right)
+ \beta_D\, \hat{\mathbf{n}} \cdot (\hat{\omega}_f \times \hat{\omega}_i),
\]
Here, $\hat{\mathbf{n}}$ is the unit vector in the direction of the final baryon’s momentum, and $\hat{\omega}_i$, $\hat{\omega}_f$ are the spin direction unit vectors of the initial and final baryons.
By choosing
\[
\hat{\omega}_f=x_f \,\hat{\omega}_i \times\hat{\mathbf{n}}+y_f \,\hat{\mathbf{n}}\times(\hat{\omega}_i\times\hat{\mathbf{n}})+z_f\,\hat{\mathbf{n}},
\]
the transition rate becomes
\begin{equation}
    R(\hat{\omega}_i,\hat{\omega}_f,\hat{\mathbf{n}})=1+\alpha_D\cos\theta+(\cos{\theta}+\alpha_D)z_f+\beta_D x_f \sin^2{\theta}+\gamma_D y_f \sin^2{\theta},
    \label{eq:transition prob}
\end{equation}
where 
\[
\cos\theta=\hat{\omega}_i\cdot\hat{\mathbf{n}}.
\]
Now, if we choose
\[
(x_f,y_f,z_f)=-\frac{1}{1+\alpha_D\cos\theta}(\beta_D,\gamma_D,\alpha_D+\cos\theta),
\]
then
\[
R(\hat{\omega_i},\hat{\omega_f},\hat{\mathbf{n}})=0.
\]
Thus, the opposite polarization direction, corresponding to an orthogonal spin state, will be the polarization direction of the daughter baryon:
\[
\hat{\omega}_f=\frac{(\alpha_D+\cos\theta)\hat{\mathbf{n}}+\beta_D\hat{\omega}_i\times\hat{\mathbf{n}}+\gamma_D\hat{\mathbf{n}}\times(\hat{\omega}_i\times\hat{\mathbf{n}})}{1+\alpha_D\cos\theta}.
\]
Alternatively, this result can also be derived by finding the extreme value of Eq.~\ref{eq:transition prob}, subject to the constraint
\[
x_f^2\sin^2\theta+y_f^2\sin^2\theta+z_f^2=1.
\]

%

\end{document}